\documentstyle[12pt]{article}
\newlength{\pubnumber} 

\catcode`\@=11
\@addtoreset{equation}{section}

\def\section{\@startsection{section}{1}{\z@}{3.5ex plus 1ex minus .2ex}
 {2.3ex plus .2ex}{\large\bf}}
\def\subsection{\@startsection{subsection}{2}{\z@}{2.3ex plus .2ex}
 {2.3ex plus .2ex}{\bf}}

\begin{document}

\def\Im{{\rm Im}}
\def\Re{{\rm Re}}
\def\gam{\gamma}
\def\tk{{\tilde k}}
\def\tal{{\tilde \al}}
\def\slc{SL(2,{\bf C})\ }
\def\ssc{\scriptscriptstyle}
\def\cm{$(c_{\ssc R}-c_{\ssc L})/2$}
\def\cR{{\cal R}}
\def\cU{{\cal U}}
\def\cA{{\cal A}}
\def\bpart{\bar\partial}
\def\part{\partial}
\def\cp{$(c_{\ssc R}+c_{\ssc L})/2$}
\def\om{\omega}
\def\ee{{\hbox{e}}}
\def\dz{{{\rm d}^2z\over 2\pi}}
\def\dD{{\rm D}}
\def\dif{{\rm d}}
\def\bsi{\bar\psi}
\def\bQ{\bar Q}
\def\bal{\bar \alpha}
\def\al{\alpha}
\def\ga{\gamma}
\def\th{\theta}
\def\bxi{\bar\xi}
\def\Gam{\Gamma}
\def\De{\Delta}
\def\de{\delta}
\def\e#1{{\rm e}^{#1}}
\def\bet{\beta}
\def\hg{{\hat g}}
\def\gh{{\hat g}}
\def\pp#1{\partial#1\bar\partial#1 - {Q_{#1}\over 4}\sqrt{\hat g}\hat R#1}
\def\ie{{\it i.e.,}\ }
\def\eg{{\it e.g.,}\ }
\def\kap{\kappa}
\def\hc{{\hat c_+}}
\def\vphi{\varphi}
\def\ups{\upsilon}
\def\vth{\vartheta}
\def\vas{\varsigma}
\def\tchi{{\tilde\chi}}
\def\kr{k_{\ssc R}}
\def\kl{k_{\ssc L}}
\def\lor#1#2{#1\leftrightarrow #2}

\def\cit#1{\cite{#1}}
\def\qq#1{Eq.~\ref{#1}}

\def\be{\begin{equation}}
\def\eeq{\end{equation}}
\def\bea{\begin{eqnarray}}
\def\eea{\end{eqnarray}}

\begin{titlepage}
\samepage{
\setcounter{page}{1}
\rightline{McGill/92-47}
\rightline{iassns-hep-92-67}
\rightline{hep-th/9210082}
\vfill

\begin{center}
 {\Large \bf Chiral non-critical strings\footnote{Talk presented
by RCM at the {\it International
Workshop on String Theory, Quantum Gravity, and the Unification
of the Fundamental Interactions}, held in Rome, Italy, 21-26 September 1992;
and at the {\it CAP/NSERC Summer
Institute on Quantum Groups, Integrable Models and Statistical Systems},
held at Queen's University, Kingston, Ontario, July 13-18, 1992.}\\}
\vfill
 {\large Robert C. Myers\footnote{rcm@hep.physics.mcgill.ca}\\}
\vspace{.25in}
{\em Physics Department, McGill University, Ernest Rutherford Building,\\
Montr\'eal, Qu\'ebec, H3A 2T8, Canada}\\
\vspace{0.3cm}
and\\
\vspace*{0.3cm}
{\large Vipul Periwal\footnote{vipul@guinness.ias.edu/vipul@iassns.bitnet}\\}
\vspace{.25in}
{\em The Institute for Advanced Study,\\
Princeton, New Jersey, 08540-4920, USA}
\end{center}
\vfill
\begin{abstract}
 {\rm It is shown that conformal matter
with $c_{\ssc L}\not=c_{\ssc R}$ can
be consistently coupled to two-dimensional `frame' gravity.
The theory is quantized in conformal gauge, following David, and Distler
and Kawai. There is no analogue of the $c=1$
barrier found in nonchiral non-critical strings.
A non-critical heterotic string is constructed---it
has 744 states in its
spectrum, transforming in the adjoint representation of $(E_8)^3.$
Correlation functions are calculated in this example, revealing
the existence of extra discrete states.}
\end{abstract}
\vfill}
\end{titlepage}

\setcounter{footnote}{0}

Conformal matter in two dimensions couples to quantum gravity via
the conformal anomaly.  The coupling is characterized by the central
charge of the conformal matter. The central charges for holomorphic
and anti-holomorphic fields in a conformal field theory may differ.
Hence, it is natural to ask how such theories,
with $c_{\ssc L}\not=c_{\ssc R},$ interact with quantum gravity.
Since these theories have
a Lorentz anomaly, in addition to the conformal anomaly, there
must be degrees of freedom other than the conformal factor
that become dynamical. Under combined scale ($\rho$) and Lorentz ($\chi$)
transformations, the zweibein transforms
$$
\ee^\pm_{\ \mu} \rightarrow\ \exp\left(\rho\pm i\chi\right)\ee^\pm_{\ \mu}\ .
$$
Just as the conformal
anomaly provides dynamics for the scale factor $\rho$,
the Lorentz anomaly provides dynamics for local
Lorentz field $\chi$. It is the action that governs these fields
that we study here.

This paper is organized as follows: Sect.~1 briefly reviews results
from the Liouville field treatment of nonchiral non-critical strings.
In sect.~2, we determine an analogous conformal
field theory representation of chiral gravity.
We consider the additional Lorentz moduli, and the
gravitational dressings in this new theory. Finally, we
derive some critical exponents.
In sect.~3 we give an example, the $\jmath$--string,
and compute the partition function and correlation functions in this example.
Sect.~4 contains some concluding remarks.

The present paper is primarily a review of the work appearing in
Ref.~\cite{cg},
and we refer the reader there for a more detailed
account of this work. Sect.~3 presents some new results about
scattering amplitudes for the $\jmath$--string.
We note that the coupling of chiral matter to two-dimensional theories
of quantum gravity has been previously considered by several
authors\cite{related,oz,tlee}. Of particular relevance to the present work
are Ref.~\cite{oz},
which studied chiral non-critical strings in
light-cone gauge, and Ref.~\cite{tlee},
which also derived the conformal field theory presented in sect.~2.

\section{Review of non-critical strings}

The initial success in producing an analytic understanding
of non-critical strings made use of light-cone gauge\cite{pol}.
The present discussion follows the David--Distler--Kawai
(DDK) approach\cite{ddk} in conformal gauge.
Scattering amplitudes in a (non-supersymmetric) string theory are
calculated with a Polyakov path integral over metrics $g_{\mu\nu}$
and matter fields $\psi$ on the two-dimensional string world-sheet
\be
\int {{\dD g}\ {\dD\psi}
\over {\hbox{vol.}(\hbox{symmetries})}}
\ \exp\!\left(-S_{o}[g,\psi]\right)\ {\cal O}_{\ssc 1}\ldots
{\cal O}_{\ssc n}\ \ .
\label{ppath}
\eeq
The classical symmetries include both diffeomorphisms and
local Weyl rescalings (\ie $g\rightarrow e^{2\rho}g$).
In the quantum theory, the path integral
measure introduces an anomaly for the latter
transformations. Polyakov\cite{geo} showed that this anomaly is proportional
to $c_m-26$, where
$c_m$ is the central charge for the matter fields.

In critical string theories, one chooses the matter fields with $c_m=26$
so that the anomaly vanishes. In this case, diffeomorphisms and Weyl
rescalings are both symmetries of the quantum
theory, and all of the local degrees of freedom of the metric decouple.
A non-critical string theory is one with $c_m\not=26$, and so the only
symmetry volume divided out in \qq{ppath} is that of diffeomorphisms.
Hence the path integral
still includes a path integral over the scale factor of the metric.
Choosing conformal gauge,  $g_{\mu\nu}=\ee^{2\rho}\hg_{\mu\nu}(m),$
and fixing diffeomorphisms \`a la Faddeev-Popov, the path integral
includes a factor of the form
\be
\int \dif m\ \dD\rho\ \exp\!\left[{c_m-26\over24\pi}\int\dif^2\!x\,
\sqrt{\hat{g}}\left\{(\hat{\nabla}\rho)^2+\hat{R}\,\rho\right\}\right]
\label{path1}
\eeq
where $\dif m$ stands for an integration over the
moduli space of the world-sheet, and $\hat{R}$ is the curvature scalar
for the background metric. The scale factor action can be regarded
as the Jacobian, which arises when the original
functional measures for the matter and
ghost fields, defined in terms of the full metric $g$, are
replaced by measures based on the background metric $\hg$.

The problem is now to understand the measure $\dD \rho,$
which is the Riemannian measure induced by
$$
(\de\rho,\de\rho) \equiv \int \dif^2x
\ \sqrt{g}\ (\de \rho)^2 = \int \dif^2x
\ e^{2\rho} \sqrt{\hg}\ (\de \rho)^2.
$$
It is much more convenient to use the translation-invariant
measure $\dD_{\ssc 0}\rho,$ induced by
$$
(\de\rho,\de\rho)_{\ssc 0} \equiv \int \dif^2x \sqrt{\hg}\ (\de \rho)^2,
$$
which allows one to treat the functional integral over the scale factor
as a standard quantum field theory. The DDK ansatz\cite{ddk} is that
\qq{path1} is replaced by
\be
\int \dif m\ \dD_{\ssc 0}\rho\ \exp\!\left[{c_m-25\over24\pi}\int\dif^2\!x\,
\sqrt{\hat{g}}\left\{(\hat{\nabla}\!\rho)^2+\hat{R}\,\rho\right\}\right]\ \ .
\label{path2}
\eeq
This result was later rigorously derived in Ref.~\cite{mmdk}.
Note that for the purposes of the present discussion, we have assumed
that a local counterterm has been introduced to produce a vanishing
cosmological constant on the world-sheet.

Given this field theory, one can perform several interesting
calculations. First, the off-diagonal components of
the stress tensor are
\bea
T_{zz} &=& -{c_m-25\over 6}\ \big(\part\rho\part\rho-\part^2\!\rho\big)\ ,
\nonumber\\
\bar T_{\bar{z}\bar{z}}
 &=& -{c_m-25\over 6}\ \big(\bpart\rho\bpart\rho-
\bpart^2\!\rho\big)\ .
\label{stresp}
\eea
The central charge computed from
\qq{stresp} is $c_\rho=26-c_m$. Therefore summing the contributions
from the scale factor, the ghosts and the matter fields, one finds that
the total central charge vanishes. This result verifies
that \qq{path2} correctly represents the integral over the world-sheet metrics,
since $c_{tot}=0$ ensures that
the total path integral is independent of the choice of the background metric
$\hg$.

Another aspect of these theories is that when a spinless primary field
$\Phi_m$ in the matter theory is inserted on the world-sheet, it
acquires a gravitational dressing $e^{\bet\rho}$. This combination produces an
operator of conformal dimension (1,1), whose position is integrated over
the surface, ${\cal O}=\int \dif^2\!x\,\sqrt{\hg}\ e^{\bet\rho}\Phi_m$.

Finally, one may calculate various
critical exponents for these theories\cite{ddk}. For instance,
the string susceptibility $\Gam$ is defined in terms of the
fixed area partition function
$$
Z(A)=\left\langle\,\delta\!\left(\int\!\dif^2x\,\sqrt{\hat{g}}\,
\ee^{\al\rho} -A\right)\,\right\rangle\propto A^{\Gam-3}\ \ .
$$
One finds the following result
\be
\Gam=2+{h-1\over12}\left[25-c_m+\sqrt{(25-c_m)(1-c_m)}\right]
\label{ggaamm1}
\eeq
where $h$ is the genus of the world-sheet. Note that $\Gam$
has a simple linear dependence on the genus $h$,
and also that it becomes complex for $c_m>1$ (and $c_m<25$).
Therefore these calculations seem to give nonsensical results in this regime.
This is the well-known
$c_m=1$ barrier. In the following sections, we will investigate
these three aspects for chiral non-critical strings:
background independence, gravitational dressings,
and critical exponents.

\section{Chiral non-critical strings}

In this section, consider chiral non-critical string theories in
which the central charges of the holomorphic and
anti-holomorphic matter fields are different. To introduce chiral
matter fields (\eg a Weyl fermion) on a curved world-sheet, one must introduce
a zweibein or frame field.
The classical symmetries are then
diffeomorphisms, Weyl scalings and local Lorentz transformations, but
anomalies arise in the quantum theory. Using a diffeomorphism-invariant
regularization, these anomalies may be described
by the non-local effective action\cite{leu}
$$
{1\over 96\pi}\int R(x) G(x,y) [(c_+-26) R(y) -ic_-U(y)]
$$
where $G$ is the inverse of the scalar Laplacian, $R$ is the curvature
scalar, and $U\equiv2\nabla\!\cdot\omega$ is given by the divergence of the
spin connection $\omega_\mu$. We also define $c_\pm \equiv
(c_{\ssc R}\pm c_{\ssc L})/2.$
An important new feature of the chiral theories is that there exists
a local counterterm, $\int \om^\mu\om_\mu$, which may be added to
this anomaly action. Therefore the coefficient of this term
will appear as a free parameter, $\xi$, in these theories,
which reflects the ambiguity in the choice of the regularization scheme.

Generically then, diffeomorphisms are the only symmetry which
survive in the quantum theory. Choosing a background gauge
$\ee^\pm_{\ \mu}=\exp(\rho\pm i\chi)\, \hat\ee^\pm_{\ \mu}$, the integration
over all zweibein  reduces to
\be
{\dD\ee\over\hbox{vol.}(\hbox{diffeo.})}=\dif m\ \dD\rho\
\dif n\ \dD\chi\ \times\ (Faddeev\!-\!Popov\ determinant)
\label{foot}
\eeq
where $\dif n$ is a integration over additional Lorentz moduli, discussed
in sect.~2.1. The functional measures for $\rho$ and $\chi$ above are
defined in terms of the full zweibein $\ee^\pm$,
but as in the nonchiral case they are exchanged for
measures based on the background zweibein $\hat{\ee}^\pm$.
This change of the two measures introduces Jacobian factors, which are in
fact identical to the factor found in nonchiral gravity.
The final functional integral over zweibein may be written as
$\int \dif m\,\dif n\,\dD_{\ssc 0}\rho\,\dD_{\ssc 0}\th\,
\ee^{-S_{\rm cft}},$ where
\be
S_{\rm cft} = \int
{\dif^2\!x\sqrt{\hat{g}}\over24\pi}\left[X
\left\{(\hat{\nabla}\!\rho)^2+\hat{R}\,\rho\right\}
 +{\xi}\left\{(\hat{\nabla}\!\th)^2
-(\hat U+{ic_-\over 2\xi}\hat R)\th\right\}\right], \label{rhoact}
\eeq
and we defined
$\th\equiv \chi-{ic_-\over2\xi}\rho,$
so that the two fields appearing in \qq{rhoact} decouple. The part of
the action for $\rho$ resembles that for the scale factor in \qq{path2}
except that the overall factor of $(c_m-25)$ is replaced by
\be
X=24-c_++\xi+{c_-^2\over4\xi} .
\label{oops}
\eeq
In \qq{rhoact}, $\th$ couples to both $\hat{R}$ and $\hat{U}$. The latter
coupling is not invariant under parity inversion on the world-sheet, and
therefore the holomorphic and anti-holomorphic components of the
stress tensor differ
\bea
T_{zz} &=& -{X\over 6}\ \left(\part\rho\part\rho-\part^2\!\rho\right)
-{\xi\over6}\ \left(\part\th\part\th+i\left({c_-\over2\xi}+1\right)
\part^2\!\th\right)\ ,
\nonumber\\
\bar T_{\bar{z}\bar{z}}
 &=& -{X\over 6}\ \left(\bpart\rho\bpart\rho-\bpart^2\!\rho\right)
-{\xi\over6}\ \left(\bpart\th\bpart\th+i\left({c_-\over2\xi}-1\right)
\bpart^2\!\th\right)\ .\nonumber
\eea
As a result the left and right central charges differ, but this produces
precisely the desired contributions, $c_{{\rm e},{\ssc R}} = 26 - c_{\ssc R}$
and $c_{{\rm e},{\ssc L}} = 26 - c_{\ssc L}$. Thus the total
central charge vanishes for both the holomorphic and anti-holomorphic
sectors, ensuring the theory is independent of the background zweibein.
Note that the free parameter $\xi$ associated with the regularization
ambiguity does not
appear in the expressions for the central charge of the combined
conformal and Lorentz induced action, but we will find that it
does affect the physical exponents.

Given a matter operator of weight $(\De_{\ssc L},\De_{\ssc R}),$ one may
expect that it acquires an exponential dressing $\exp[\al\rho+ik\th]$
to make it a (1,1) operator, as in nonchiral gravity\cit{ddk}.
Note that this is possible even when $\De_{\ssc L}\not=\De_{\ssc R}$,
because $\exp[ik\th]$ has different left and right weights due to the form
of the stress tensor. Explicitly, one finds $k=\De_{\ssc L}-
\De_{\ssc R}$ for such a dressing. In fact, this description is incomplete
as we will find in the next subsection.

\subsection{Lorentz moduli}

In \qq{foot}, the integration over all frames includes an ordinary integral
over Lorentz moduli $dn$. These moduli are extra global phases
that must be integrated over,
above and beyond the moduli associated with the integration over
conformal equivalence classes of metrics.
One can realize these phases by shifting the spin connection,
$\om\rightarrow\om+\sum_{i=1}^{2h}\lambda_i\,\beta^i$, where $\beta^i$ is a
basis for the harmonic differentials on the genus $h$ surface. Since
$\beta^i$ are closed and divergenceless, this shift leaves $R$
and $U$ everywhere unchanged.
Given any closed contour around a particular
nontrivial cycle though, one acquires an additional phase:
$\int_a\om\rightarrow\int_a\om+\lambda_a$.
These global phases are most conveniently incorporated into the Lorentz field
$\th$ (rather than the background geometry). To be precise, one lets
$$
\dif\th=\dif\tilde{\th}+\sum_{i=1}^{2h}\lambda_i\,\beta^i\ .
$$
where $\tilde{\th}$ is a (single-valued) function on the surface.
Since the $\beta^i$
are not exact forms, $\th$ must be multivalued on the surface
(or alternatively, $\th$ contains discontinuities).
The measure for the Lorentz moduli,
$\dif n=\prod\dif\lambda_i$, would then be
included as a part of the functional measure $\dD_{\ssc 0}\th$.

These global phases are associated with the nontrivial cycles on
higher genus surfaces. Additional nontrivial cycles occur  in correlation
functions surrounding the operator insertions. Therefore we should
allow $\th$ to have discontinuities around such cycles.
Such cuts would be produced by dressing the matter operators with
exponentials of the form: $\exp[\al\rho+i\kr\th_{\ssc R}+i\kl\th_{\ssc L}]$.
Now given a matter operator of weight $(\De_{\ssc L},\De_{\ssc R}),$ one may
fix $\al$ and $\kr+\kl$ in the gravitational dressing
to produce a (1,1) operator. This leaves one free parameter, namely
$\kr-\kl$. Hence in chiral
gravity, each matter operator acquires a family of gravitational dressings,
corresponding to a continuum of Lorentz phases on the contours
enclosing the operator.

These dressing operators are nonlocal, and hence may seem unnatural.
In fact, one is forced to introduce such operators because of another
aspect of the theory. Because $\rho$ and $\theta$ couple to background
charges, their momentum conservation rules are nontrivial.
One finds the following superselection rules
\bea
\sum_{a} \al^{\ssc (a)} &=& {X\over3} (1-h), \nonumber\\
\sum_{a} k^{\ssc (a)}_{\ssc R} &=& -{{\xi}\over 3}\left[{c_-\over2\xi}+1
\right] (1-h), \nonumber\\
\sum_{\ssc a} k^{\ssc (a)}_{\ssc L} &=& -{{\xi}\over 3}\left[{c_-\over2\xi}-1
\right](1-h). \label{ssrho}
\eea
These results may be derived by demanding \slc invariance of scattering
amplitudes on the sphere. It is the novel coupling of $\th$ to
the divergence of the background spin connection $\hat{U}$, which leads
to $\sum\kr\not=\sum\kl$. As a result, nonvanishing amplitudes must include
nonlocal operators of the form discussed above.

\subsection{String susceptibility}

In this section, we examine critical exponents to address the question of the
critical `dimension' for chiral gravity. One might begin by demanding
the reality of the string susceptibility\cit{ddk}. Define the area operator,
$\int\!\dif^2x\,\sqrt{\hat{g}}\,\ee^{\al\rho}$ with $\alpha=
[X-\sqrt{X(X-24)}]/6$. This definition is chosen to involve only
a local dressing operator,
and to make no explicit reference to the Lorentz field, which should
be irrelevant to defining the area in analogy to the classical geometry.
Then one considers the fixed area partition function
$$
Z(A)=\left\langle\,\delta\!\left(\int\!\dif^2x\,\sqrt{\hat{g}}\,
\ee^{\al\rho} -A\right)\,\right\rangle=0\ \ .
$$
The result vanishes, except for genus one
surfaces,\footnote{For $h=1$,
one finds $Z(A)\propto A^{\Gamma-3}$ with $\Gamma=2$, exactly as in
nonchiral gravity\cit{ddk}.} because the super-selection rules for
$\th$ in \qq{ssrho} are not satisfied.  To properly fix the
{\it two} $\th$ zero mode
integrals, one can consider inserting punctures with dressings which
absorb the appropriate $\th$-momenta. If one introduces a single puncture,
there is a unique dressing which yields a nonvanishing
result for $g=0,1$. $P_{\ssc 0}
=\int\!\dif^2x\sqrt{\hat{g}}\exp[\beta\rho+i\kr
\th_{\ssc R}+i{\kl}\th_{\ssc L}]$ with $\beta=\al$ precisely
as in the area operator, and $\kr=(h-1)(c_-/6+\xi/3)$
and ${\kl}=(h-1)(c_-/6-\xi/3)$. Now one has
$$
Z'(A)=\left\langle\,P_{\ssc 0}
\ \delta\!\left(\int\!\dif^2x\,\sqrt{\hat{g}}\,\ee^{\al\rho}
-A\right)\,\right\rangle\propto A^{\Gamma-2}
$$
where
\be
\Gamma=2+{h-1\over12}\left[X+\sqrt{X(X-24)}\right]
\label{ggaamm}
\eeq
for $h=0,1.$
This is precisely analogous to the result for nonchiral gravity\cit{ddk}
with the replacement $(25-c_m)\rightarrow X$.
If we make the assumption that $X$ is positive so that the $\rho$-action
\qq{rhoact} has a positive coefficient, then from \qq{oops}
requiring that $\Gamma$ be real imposes the restriction
\be
{X-24}\ \ =\ \ \xi+{c_-^2\over4\xi}-c_+>0\ \ .\label{jacket}
\eeq
For any fixed values of $c_{\pm}$, there will exist values of $\xi$
for which this inequality is satisfied. Therefore chiral gravity has no
barriers for the allowed values of central charges in the matter sector.

Of course, this conclusion relies on results for only $h=0,1,$ since
$Z'(A)$ always vanishes for $h>1$. Introducing two punctures yields nontrivial
results for higher genera as well. In this case, one finds
that $Z''(A)\propto A^{{\widetilde\Gamma}-1}$ where $\widetilde\Gamma
=\Gamma+(\al_++\al_--2\al)/\al$. $\Gamma$ is given by
\qq{ggaamm}, while
$\al_\pm=\left[X+\sqrt{XY_\pm}\right]/6$ with
$$
Y_\pm=\left[2h^2-1\pm2h\sqrt{h^2-1}\,\right]\xi+
        \left[2h^2-1\mp2h\sqrt{h^2-1}\,\right]{c_-^2\over4\xi}-c_+
$$
for $h\ge1$. Thus one finds that the genus dependence of the string
susceptibility is far more complicated than the simple linear dependence
found in \qq{ggaamm1} for nonchiral gravity. Further requiring real exponents
by imposing $Y_\pm>0$, only produces constraints which are less restrictive
than \qq{jacket}. Finally note that we have ignored ghost zero
modes in this entire discussion, but that a more rigorous account can be
given completely equivalent to that appearing in Ref.~\cite{ddk}.

\section{The $\jmath$--string}

We now consider as a concrete example the heterotic non-critical string
theory constructed from the holomorphic conformal field theory
associated with the
$E_{\ssc 8}\times E_{\ssc 8}\times E_{\ssc 8}$ root lattice.  One may
view the world-sheet matter as 24 chiral bosons, or as 48 chiral fermions
with appropriate sums over spin structures. We designate this theory
as the $\jmath$--string for the amusing result that the partition function
is the $\jmath$--invariant\cite{jps} on the torus.

The calculation of the torus partition function for this theory is
straightforward\cite{dp}. The only novel feature is the
$\th$ factor which incorporates the Lorentz
moduli. Covering the torus with a fixed coordinate patch,
$0\le\sigma^{\ssc1},\sigma^{\ssc2}\le1$, the world-sheet metric may be
written
$\dif s^2=|\dif\sigma^{\ssc1}+\tau\,\dif\sigma^{\ssc2}|^2$.
We introduce arbitrary phases in $\th$
around the $\sigma^i$ cycles by setting
$$
\th(\sigma^{\ssc1},\sigma^{\ssc2})
={\tilde\th}(\sigma^{\ssc1},\sigma^{\ssc2})+2\pi\lambda_i\sigma^i.
$$
Here ${\tilde\th}$ is a smooth function on the torus, and the second term
incorporates the multivalued contribution of the Lorentz moduli.
The contribution to the partition function is then found to be
\bea
Z_\th&=&(\Im\tau)^{-1/2}|\eta(q)|^{-2}\int_{-\infty}^{\infty}
\dif\lambda_{\ssc1}\dif\lambda_{\ssc2}
\ \exp\left(-{\pi\xi|\lambda_{\ssc2}-\lambda_{\ssc1} %
\tau|^2\over6\,\Im\tau}\right)
\nonumber\\
&=&(\Im\tau)^{-1/2}|\eta(q)|^{-2}\ {6/\xi}
\label{lorry}
\eea
where $q\equiv \exp(2\pi i\tau)$, and
we have included the phase integral, which one may note is
invariant under modular transformations.
After the phase integral is performed, the final $\th$ factor
is simply that of a free scalar field (up to an arbitrary normalization).

Combining Eq.~\ref{lorry} with the contributions of the other fields yields
$$
Z_{\ssc\rm torus} = \int {\dif^2\tau\over(\Im\tau)^2}\ \jmath(\tau)
= \int {\dif^2\tau\over(\Im\tau)^2}
\ \left({\Theta_2(q)^2+\Theta_3(q)^2+\Theta_4(q)^2\over\eta(q)^8}\right)^3\ .
$$
To understand the spectrum
of the theory, we investigate the region of
large $\Im\tau,$ where
\be
\jmath(\tau)\sim {1\over q} + 744 +\dots \qquad.
\label{744}
\eeq
The integration over $\Re\tau$ projects out any term in Eq.~\ref{744} with
a non-zero power of $q.$
The only term left is the constant $744,$ indicating that there are
$744$ states in this string theory. Of these, $720$ correspond to
winding number states in the $24$--dimensional lattice, and $24$ come
from the maximal torus of the group, the so-called oscillator states.
If one counted the states before performing the phase integral
in Eq.~\ref{lorry},
each of the 744 states corresponds to a continuum with arbitrary phases
around the $\sigma^{\ssc1}$ cycle.

\subsection{Correlation functions}

We now consider some sample correlation functions in this theory.
In the partition function, we found that the physical states
correspond to (0,1) operators in the matter sector. Thus we
construct exponential (1,0) dressings, as discussed in sect.~2.1.
First though in the present theory with $c_\pm=12$, computations are
greatly simplified if one
rescales the fields to a new basis ($\hat\rho=(x+2/x)\rho,\,\hat\th=x\th$)
where $x=\sqrt{\xi/3}$.
There are two possible (1,0) operators of the form
$\exp[\al\hat\rho+ik\hat\th_{\ssc R}+i\tk\hat\th_{\ssc L}]$,
with good semiclassical limits. For fixed $k$, they correspond to
\begin{itemize}
\setlength{\itemsep}{0pt}
\item[i.] $\tk=k+x,\ \al=-k$
\item[ii.] $\tk=-2/x-k,\ \al=-k\ \ .$
\end{itemize}
\par\noindent
The first introduces a fixed phase in the $\th$ field with
$k-\tk=-x$, while the phase of the second dressing varies continuously
with $k$, $k-\tk=2/x+2k$. The momentum super-selection rules for correlation
functions are: $\sum\al^{\ssc (a)}=x+2/x=-\sum k^{\ssc (a)}$,
and $\sum \tk^{\ssc (a)}=x-2/x$.

One can begin by considering three-point amplitudes of some given
set of matter operators. For a fixed set, there would be eight
possible amplitudes corresponding to all of the different combinations
of dressings. Some of these amplitudes vanish since they combine
dressings which are incompatible with the momentum conservation rules.
The amplitudes which survive are those which involve either
two (i) and one (ii) dressings, or one (i) and two (ii).
In those cases where the amplitude is non-vanishing, one
finds that as expected the results are \slc invariant, but
also {\it independent} of
how the various dressings are combined with the matter operators.
There are two classes of nonvanishing amplitudes: those with three
winding number states, $\exp(i\gamma^{\ssc(a)}\!\cdot\! X_{\ssc R})$,
which yield
simply $\delta(\gamma^{\ssc(1)}+\gamma^{\ssc(2)}+\gamma^{\ssc(3)})$,
and those with two winding number states and one oscillator state,
$i\beta\!\cdot\!\part X_{\ssc R}$, which yield
$\beta\cdot\gamma^{\ssc(1)}\ \delta(\gamma^{\ssc(1)}
+\gamma^{\ssc(2)})$.\footnote{These results use %
Kronecker $\delta$-functions, since the winding number vectors are discrete.}

The four-point amplitudes produce more interesting results, as we
illustrate here. Consider the following correlation function
\be
\cA=\int\dif^2z^{\ssc(3)}\ \Big\langle\
c\bar c\, V_{\rm i}\, \ee^{i\gamma^{\ssc(1)}\cdot X_{\ssc R}}(z^{\ssc(1)})
\ c\bar c\, V_{\rm i}\, \ee^{i\gamma^{\ssc(2)}\cdot X_{\ssc R}}(z^{\ssc(2)})
\ V_{\rm ii}\, \ee^{i\gamma^{\ssc(3)}\cdot X_{\ssc R}}(z^{\ssc(3)})
\ c\bar c\, V_{\rm i}\, \ee^{i\gamma^{\ssc(4)}\cdot
X_{\ssc R}}(z^{\ssc(4)})\ \Big\rangle\ ,
\label{amplo}
\eeq
where $c\bar c$ are ghost dressings for the fixed operators, $V_{\rm i(ii)}$
are gravitational dressings of type i (ii) given above, and
$\exp[i\gamma^{\ssc(a)}\cdot X_{\ssc R}]$ are winding state operators in
the matter sector with $\gamma^{\ssc(a)}\cdot\gamma^{\ssc(a)}=2$.
Momentum conservation requires $\sum\gamma^{\ssc(a)}=0$ in the matter
sector, while the super-selection rules, for the gravity sector
given above, restrict the momenta to be
\begin{itemize}
\setlength{\itemsep}{0pt}
\item[$z^{\ssc(1)}$:] $k=q$, $\tk=q+x$, $\al=-q$
\item[$z^{\ssc(2)}$:] $k=p$, $\tk=p+x$, $\al=-p$
\item[$z^{\ssc(3)}$:] $k=x/2-1/x$, $\tk=-1/x-x/2$, $\al=-x/2+1/x$
\item[$z^{\ssc(4)}$:] $k=-3x/2-1/x-p-q$, $\tk=-x/2-1/x-p-q$,
$\al=3x/2+1/x+p+q\ .$
\end{itemize}
\par\noindent
One can explicitly verify that the result is \slc invariant,
and fixing $z^{\ssc(a)}=(\infty,1,z,0)$ yields
\be
\cA=\int\dif^2z\ (1-z)^{\gamma^{\ssc(2)}\cdot\gamma^{\ssc(3)}}
\ z^{\gamma^{\ssc(4)}\cdot\gamma^{\ssc(3)}}
\ (1-\bar z)^{-1-x^2/2-px}\ {\bar z}^{x^2+x(p+q)}\ \ .
\label{ample}
\eeq
On the holomorphic side of this integral, one always has integral
exponents since $\gamma^{\ssc(a)}\cdot\gamma^{\ssc(b)}=-2,-1,0,1,2$,
while on the anti-holomorphic side, arbitrary exponents arise as $p$
and $q$ are varied.

Thus we would like to evaluate integrals of the form
\be
\cA (a,b,c,d)=\int\dif^2z\ (1-z)^a\ z^b\ (1-\bar z)^c\ {\bar z}^d\ \ .
\label{hoot}
\eeq
At present, we do not know how to perform this integral for arbitrary
exponents, but it is straightforward to evaluate in the special case when
several of the exponents are integers. Using repeated
integration by parts and the identity, $\bar{\part}(z-w)^{-1}=\pi
\de^2(z-w)$, one finds
\be
\cA (a,b,c,d)=\pi{\Gam(-a-b-1)\over\Gam(-a)\Gam(-b)}{\Gam(c+1)\Gam(d+1)\over
\Gam(c+d+2)}
\label{geewhiz}
\eeq
which is valid when $a+b+c+d<-2$, $a$ and $b$ are negative integers, and
$c$ is a positive integer while $d>-1$ or $d$ is a positive integer
while $c>-1$. Eq.~\ref{geewhiz} is an enticing formula since setting
$a=c$ and $b=d$ precisely reproduces the Virasoro-Shapiro amplitude\cite{vs},
as $\cA(a,b,a,b)$ should. If we allow the exponents $c$ and $d$
to take arbitrary values in this result (as they would in Eq.~\ref{ample}),
we find that the amplitude
has an infinite set of poles for negative integer values of $c$ and $d$.
This would be a curious result since the partition function indicated that the
theory only contains a finite number of physical states.
One can disregard these poles though since we have not justified
\qq{geewhiz} as the correct analytic continuation
for \qq{hoot} with arbitrary exponents.

Even within the restrictions for the derivation for Eq.~\ref{geewhiz},
we do in fact find that $\cA$ has unexpected poles. We illustrate this
by focusing on one particular choice of exponents:
$a=\gamma^{\ssc(2)}\cdot\gamma^{\ssc(3)}=-2$,
$b=\gamma^{\ssc(4)}\cdot\gamma^{\ssc(3)}=-2$, and
$d=x^2+x(p+q)=2$. In this case, Eq.~\ref{geewhiz} reduces to
\be
\cA={4\pi\over(c+1)(c+2)(c+3)}
\label{poles}
\eeq
where $c=-1-x^2/2-px$. While the original derivation
applies for $-1<c<0$, \qq{poles} provides a unique analytic continuation to the
entire complex plane, with poles at $c=-1,-2,-3.$ In terms of the
$\th_{\ssc R}$-momenta appearing in the external states
in \qq{amplo}, these poles
occur at $p=-{x\over2}+{n\over x}$ and $q=-{x\over2}+{2-n\over x}$ where
$n=0,1,2$.

Now the momentum super-selection rules may be used
to determine the character of the intermediate states producing these poles.
For instance, $\al^{\ssc (a)}=-k^{\ssc (a)}$ is satisfied by
all of the external states, and through the momentum conservation
rules, must also be satisfied by the intermediate states. As a result, the
dressing for the latter states has $\De_{\ssc R}=0$, and so they
will correspond to (0,1) matter operators, as expected.\footnote{The
conclusions, here and below, with regard to the dressings are unaffected
if the arguments are extended for the possible inclusion of unconventional
ghost dressings.} Since $a$, $b$ and $d$ are fixed, one must first
determine in what channel the poles occur.
Note that since
$\gamma^{\ssc(a)}\cdot\gamma^{\ssc(a)}=2$ and
$\sum \gamma^{\ssc(a)}=0$ with $\gamma^{\ssc(2)}\cdot\gamma^{\ssc(3)}=-2=
\gamma^{\ssc(4)}\cdot\gamma^{\ssc(3)}$, one must have
$\gamma^{\ssc(1)}=-\gamma^{\ssc(2)}
=\gamma^{\ssc(3)}=-\gamma^{\ssc(4)}$. Therefore $\cA$ cannot factorize
in the (3,1) particle
channel but only in the (3,2) or (3,4) channels where
the external winding number states can couple to an oscillator state
in the intermediate channel. Examining the (3,2) channel, one finds that
the exponential dressing would have a fixed weight $\De_{\ssc L}=4$
at all of the poles. Since this is not one(, zero or a negative
integer---see below), the amplitude cannot factorize in this
channel as two \slc invariant three-point amplitudes.

Hence one is
lead to conclude that the poles must occur in the (3,4)
channel, for which one finds the exponential dressings have
$\De_{\ssc L}=1-n$. For $n=0$ (\ie $c=-1$), the intermediate state
is a conventional state combining a (0,1) matter operator,
$i\beta\!\cdot\!\part X_{\ssc R}$, with an exponential ($\hat{\rho}$,
$\hat{\th}$)
dressing with weight (1,0). However for $n=1,2$ (\ie $c=-2,-3$),
the exponential gravity dressings have $\De_{\ssc L}=0,-1$.
Thus the complete dressings cannot be simply exponentials, but must also
have oscillator contributions to produce $\De_{\ssc L}=1$. Hence, the
intermediate states must involve gravity dressings of the form:
$(A\,\bar{\part}\hat{\rho}+B\,\bar{\part}\hat{\th})
\exp[\al\hat\rho+ik\hat\th_{\ssc R}+i\tk\hat\th_{\ssc L}]$ or
$(A\,\bar{\part}^2\!\hat{\rho}+B\,\bar{\part}^2\!\hat{\th}+C(\bar{\part}
\hat{\rho})^2+D
(\bar{\part}\hat{\th})^2+E\,\bar{\part}\hat{\rho}\,\bar{\part}\hat{\th})
\exp[\al\hat\rho+ik\hat\th_{\ssc R}+i\tk\hat\th_{\ssc L}]$.
Note that it is possible to construct dressings with
oscillator contributions in chiral gravity, because the gravity sector
includes two scalar fields. This contrasts with the nonchiral case,
which only involves
the Liouville field, and hence only allows for exponential dressings.

In fact the above description is still incomplete. Explicit construction
of a (1,0) primary field
of the form $(A\,\bar{\part}\hat{\rho}+B\,\bar{\part}\hat{\th})
\exp[\al\hat\rho+ik\hat\th_{\ssc R}+i\tk\hat\th_{\ssc L}]$ produces
a unique solution up to an overall normalization:
$\bar{\part}(
\exp[\al\hat\rho+ik\hat\th_{\ssc R}+i\tk\hat\th_{\ssc L}])$, where the
momenta are chosen such that the exponential is a (0,0) operator. It is
easy to verify that states with such a total derivative dressing decouple.
Therefore the correct dressing for the intermediate state at the $c=-2$
pole must incorporate the ghost number current as well,
$(A\bar{\part}\rho+B\bar{\part}\th+C\bar{c}\bar{b})
\exp[\al\hat\rho+ik\hat\th_{\ssc R}+i\tk\hat\th_{\ssc L}]$.
In this respect, the states producing these extra poles resemble
the discrete states appearing in the $D=1$ non-critical string\cite{disc}.
Another feature in common with the latter is that these new
states make no appearance in the analysis of the partition function
presented above\cite{parti}.

\section{Concluding remarks}

We have examined two-dimensional quantum gravity coupled to chiral
matter, and found that for fixed values of $c_\pm$, there is in fact a
family of theories labelled by the free parameter, $\xi$. This freedom
reflects an ambiguity in the choice of a diffeomorphism invariant
regularization scheme used to define the theory.
The effects of $\xi$ are quite intricate:
It does not
appear in the central charge of the combined
conformal and Lorentz induced action, nor in the
partition function.
However, $\xi$ does affect critical exponents,
the spectrum of discrete states, and
the positions of poles in amplitudes. One might
think of the freedom to specify $\xi$
in analogy with the cosmological constant,
which arises as a free parameter in nonchiral gravity.
In fact since chiral gravity
contains two scalar fields, it is possible to construct a large
number of new (1,1) primary fields, beyond the cosmological
constant operator, which may be included as terms in the action.
Then, $\xi$ should be counted as only one of the undetermined
couplings which arise in association with these terms.

A physically consistent
analysis required some restrictions on $\xi$ in the form of the
inequality \qq{jacket}, but no barriers appeared for the matter theories.
Note as well that one can easily produce theories with
space-time tachyons, which do not yield unphysical critical exponents.
Thus at this level, the properties of space-time tachyons and
physical consistency on the world-sheet,
appear to be divorced in the present theory, in contrast to
nonchiral gravity\cit{seib2}. Clearly, the appearance of the Lorentz
field has drastic effects on the quantum theory of the world-sheet
geometry. The most pressing question would appear to be to
understand the complete space of physical states\cit{next}.

\bigskip
R.C.M. was supported by NSERC of Canada, and Fonds
FCAR du Qu\'ebec. V.P. was supported by D.O.E. grant
DE-FG02-90ER40542.

\bibliographystyle{unsrt}

\begin{thebibliography}{99}

\bibitem{cg} R.C. Myers and V. Periwal, ``Chiral gravity in two dimensions,''
hep-th/9207117, to appear in {\em Nuclear Physics B}.

\bibitem{related} K. Li, {\em Phys. Rev.} {\bf D34} (1986) 2292;
T. Fukuyama and K. Kamimura, {\em Phys. Lett.} {\bf 200B} (1988) 75;
T. Berger and I. Tsutsui, {\em Nucl. Phys.} {\bf B335} (1990)245;
{\em Z. Phys.} {\bf C49} (1991) 337; S. Hwang and R. Marnelius,
{\em Nucl. Phys.} {\bf B271} (1986) 369.

\bibitem{oz} Y. Oz, J. Pawelczyk and S. Yankielowicz, {\em Phys. Lett.}
{\bf B249} (1990) 417; {\em Nucl. Phys.} {\bf B363} (1991) 555.

\bibitem{tlee} T. Lee, `Two-Dimensional Chiral Quantum Gravity,'
preprint SNUTP-91-54.

\bibitem{pol} A.M. Polyakov, {\em Mod. Phys. Lett.} {\bf A2} (1987) 893;
V.G. Knizhnik, A.M. Polyakov and A.B. Zamolodchikov, {\em
Mod. Phys. Lett.} {\bf A3} (1988) 819.

\bibitem{ddk} F. David, {\em Mod. Phys. Lett.} {\bf A3} (1988) 1651;
J. Distler and H. Kawai, {\em Nucl. Phys.} {\bf B321} (1989) 509.

\bibitem{geo} A.M. Polyakov, {\em Phys. Rev. Lett.} {\bf 103B} (1981) 207.

\bibitem{mmdk} N.E. Mavromatos and J.L. Miramontes, {\em Mod. Phys. Lett.}
{\bf A4} (1989) 1847; E. D'Hoker and P.S. Kurzepa, {\em Mod. Phys. Lett.}
{\bf A5} (1990) 1411;
E. D'Hoker, {\em Mod. Phys. Lett.} {\bf A6} (1991) 745.

\bibitem{leu} H. Leutwyler, {\em Phys. Lett.} {\bf 153B} (1985) 65.

\bibitem{jps} J.-P.~Serre, {\em A Course in Arithmetic}, (Springer, New
York, 1973).

\bibitem{dp} See, for example,
E.~D'Hoker and D.H.~Phong, {\em Rev. Mod. Phys.} {\bf 60}
(1988) 917.

\bibitem{vs} M.A. Virasoro, {\em Phys. Rev.} {\bf 177} (1969) 2309;
J. Shapiro, {\em Phys. Rev.} {\bf 179} (1969) 1345.

\bibitem{disc} D.J. Gross, I.R. Klebanov and M. Newman, {\em Nucl.
Phys.} {\bf B350} (1991) 621; A.M. Polyakov, {\em Mod. Phys. Lett.}
{\bf A6} (1991) 635; ``Singular States in 2D Quantum Gravity,''
preprint PUPT-1289 (1991).

\bibitem{parti} M. Bershadsky and I.R. Klebanov, {\em Phys. Rev. Lett.}
{\bf 65} (1990) 3088; {\em Nucl. Phys.} {\bf B360} (1991) 559.

\bibitem{seib2} N. Seiberg, {\em Prog. Theor. Phys. Suppl.}
{\bf 102} (1990) 319.

\bibitem{next} In preparation

\end{thebibliography}

\end{document}